\documentclass[aps,prl,twocolumn,showpacs,superscriptaddress,groupedaddress]{revtex4} 
\usepackage{amsmath,amssymb,amsfonts,amsthm,amscd, color}
\usepackage{enumerate}
\usepackage[hang, flushmargin]{footmisc}
\usepackage{natbib}
\usepackage{enumerate}
\usepackage{graphicx}
\usepackage{hyperref}

\def\be{\begin{equation}}
\def\ee{\end{equation}}
\def\ba{\begin{eqnarray}}
\def\ea{\end{eqnarray}}

\def\L{\mathcal{L}}

\def\f{\frac}

\def\H{\mathcal{H}}

\def\E{\mathcal{E}}
\def\P{\pi}

\begin{document} 

\title{Volume Weighting the Measure of the Universe from Classical Slow-Roll Expansion}
\author{David Sloan}
\email{david.sloan@physics.ox.ac.uk}
\affiliation{Beecroft Institute of Particle Astrophysics and Cosmology, Department of Physics,
University of Oxford, Denys Wilkinson Building, 1 Keble Road, Oxford, OX1 3RH, UK}
\author{Joseph Silk}
\email{silk@iap.fr}
\affiliation{Beecroft Institute of Particle Astrophysics and Cosmology, Department of Physics,
University of Oxford, Denys Wilkinson Building, 1 Keble Road, Oxford, OX1 3RH, UK}
\affiliation{Institut d'Astrophysique de Paris, UMR 7095, CNRS, UPMC Universit\'{e} Paris 6, Sorbonne Universit\'{e}s, 98 bis boulevard Arago, 75014 Paris, France}
\affiliation{Laboratoire AIM-Paris-Saclay, CEA/DSM/IRFU, CNRS, Universite Paris Diderot, F-91191 Gif-sur-Yvette, France}
\affiliation{Department of Physics \& Astronomy, The Johns Hopkins University, \\
3400 N Charles Street, Baltimore, MD 21218, USA}

\begin{abstract}

One of the most frustrating issues in early universe cosmology centers on how to reconcile the vast choice of universes in string theory and in its most plausible high energy sibling, eternal inflation, that jointly generate the string landscape with the fine-tuned and hence relatively small number of universes that have undergone a large expansion and can accommodate observers and, in particular, galaxies. We show that such observations are highly favored for any system whereby physical parameters are distributed at a high energy scale, due to the conservation of the Liouville measure and the gauge nature of volume, asymptotically approaching a period of large isotropic expansion characterised by $w=-1$. Our interpretation predicts that all observational probes for deviations from $w=-1$ in the foreseeable future are doomed to failure. The purpose of this paper is not to introduce a new measure for the multiverse, but rather to show how what is perhaps the most natural and well known measure, volume weighting, arises as a consequence of the conservation of the Liouville measure on phase space during the classical slow-roll expansion.

\end{abstract}

\maketitle
Cosmology in its standard form, the so-called $\Lambda CDM$ model, provides a remarkably successful fit to all current data. The latest developments, from the large-scale structure of the galaxy distribution, dark matter surveys via gravitational lensing, dark energy via supernovae and cosmic microwave background fluctuations via orbital and suborbital telescopes, have ushered in the age of precision cosmology. Inflationary cosmology has been especially adept in pushing the frontier of our knowledge of the initial conditions back to the epoch of quantum gravity, the Planck epoch, where one enters the intellectually appealing domain of string theory. However this has occurred at a severe cost, namely that of introducing the landscape of some $10^{500}$ Calabi-Yau manifolds, any one of which is equally likely to contain our present universe in the absence of any selection principle. This is a potential disaster, since galaxies could not form in almost all of these future universes, and has led to the introduction of anthropic reasoning as a means of selecting a universe with cosmological constant comparable to the observed value \cite{Weinberg:1987dv}, observed to be exceedingly small in terms of fundamental units from a set where the typical value is larger by 120 orders of magnitude.

For many physicists, this is not a satisfactory resolution. More significantly, in the absence of a measure that enables us to assess the probability of any given universe in the multiverse, the problem is not even well posed. There have been numerous attempts to define the measure of the universe, but to date, none have been completely successful. Moreover, many of these measure prescriptions rely on implementation via anthropic reasoning. This is equally true in string theory \cite{gibbons:2006pa, Vilenkin:1994ua}, eternal inflation \cite{ Winitzki:2008yb,Linde:2010xz, Guth:2011ie} and loop quantum gravity \cite{Corichi:2010zp,Measure2, Corichi:2013kua, Measure}.

Here we propose a novel solution that accounts for the size and even acceleration of the universe. Our idea centers on volume-weighting, and has been discussed previously \cite{Hawking:2007vf,Winitzki:2008yb}. However we show here that it arises naturally as a result of a dynamical systems approach to cosmology. Our goal is to show that the volume weighting is a direct consequence of the conservation of the symplectic structure under Hamiltonian flow during the classical expansion of the universe. Volume weighting has been examined extensively in the context of eternal inflation, sometimes referred to as a `Stationary Measure'. In \cite{Linde:2007nm} it was shown that, in the case of simple models at least, such a measure produced gauge invariant results - a key issue plaguing many other proposals such as time or scale factor cut-offs. In particular, given a set of possible de-Sitter vacua, the probability of an observer finding themselves in such a vacuum is proportional to the amount of volume growth during the slow-roll expansion of the vacuum. In \cite{Linde:2008xf} this proposal was shown to be free of both the youngness paradox (that is there is not a high likelihood that our bubble was very recently created) and of Boltzmann brains. The analysis we present here is complementary to such approaches; from the perspective of eternal inflation, we will examine the behaviour of a single bubble and the evolution of a measure during the slow-roll inflationary phase. 

The physical observables defining the universe are determined according to some unknown probability distribution at an initial energy density. This is motivated by the fact that general relativity coupled to the standard model is an excellent fit to observational evidence at low energies in the classical regime. We consider the probability of an observer at low energy density making physical measurements which are compatible with a given evolution of the universe. The strategy is the following: we express general relativity (GR) coupled to matter in a Hamiltonian setting and separate phase-space variables into those which form physical observables and those which are gauge. We take the probability distribution on physical observables, expressed in terms of non-gauge variables, and `thicken' this to a form which is proportional to the Liouville measure. We do not base our measure on the Liouville measure itself, but rather observe that any measure formed from physical observables can be thickened to one proportional to the Liouville measure by taking exterior products with gauge directions to cover the whole of phase space. This measure is then conserved through evolution, and we can thus project back down by integrating out the gauge directions to form a new probability distribution at late times to describe the observations made at low energy. Thus we express the induced low density probability distribution in terms of the high density distribution, using Liouville's theorem to translate between the two. Critically, the length of the interval in gauge direction may not be conserved under evolution, thus on projecting back to physical observables the interval in this gauge direction must be adapted to cover the same set of solutions as chosen at the high energy surface. This induces a weighting on the low energy probability distribution proportional to the relative expansion of gauge directions.

In \cite{gibbons:2006pa}, the Liouville measure was directly applied at a low energy density to claim that inflation was an improbable event. The discrepancies between this and the corresponding result for high energy was explored in \cite{Corichi:2010zp, Measure2}, and the resulting interpretation as an implicit probability weighting given in \cite{Corichi:2013kua}. Our approach differs in that we allow for any choice of measure at high energy and show that when translated to observations at low energy a re-weighting based on increase in volume between equal energy slices arises naturally. This differs significantly from volume-weighting directly \cite{Winitzki:2008yb} which is introduced by hand.

We will consider an action consisting of matter satisfying a Lagrangian $\L_{matter}$ minimally coupled to gravity on a topology that splits into a spatial slice $\Sigma$ and a time-like direction. The matter Lagrangian will be kept general - we shall only require that it obeys the weak energy conditions. A more inflationary focussed exploration of such dynamical systems is given in \cite{Alho:2015cza}. For brevity of exposition, we will take gravity to be described by GR, however a wide class of gravitational theories exhibits qualititatively similar properties \cite{MinCoup}, and we shall assume $\Sigma$ to be flat and compact, although these assumptions can be relaxed in appropriate limits. 

\be \label{action} S=\int_I dt \int_{\Sigma} \sqrt{g}(R - \L_{matter}) \ee

We shall assume that our manifold admits a constant mean curvature foliation \cite{Rendall} . This is not highly restrictive, and largely consists of assuming weak energy conditions on our matter Lagrangian. Thus the mean curvature (Hubble parameter) $H$ will be monotonically non-increasing between slices and we can reparametrize our time interval as an interval in $H$. Since we are concerned with properties observed at a given energy density, or equivalently, Hubble parameter, we identify a dynamical trajectory as a single entity which is determined by the values of physical parameters at a given $H$, from the set of all trajectories which obey the Hamiltonian constraint. 

This choice of time foliation by Hubble parameter is presented for reasons of brevity, and is not strictly necessary. From a physical perspective, any monotonic parameter can be chosen to form a good clock at least in the classical regime. During the highly quantum mechanical phase of eternal inflation, the Hubble parameter fails to satisfy this condition, but a different choice of time foliation can be employed - the standard route being the use of a massless scalar field minimally coupled to gravity. Following such a procedure produces an analogous result to that given here. 

The action carries naturally a two-form on $\Sigma$, the symplectic structure $\omega$, which is conserved between Cauchy slices. We further may decompose our fields into eigenfunctions $E_i$ of the Laplace operator compatible with $\Sigma$ (Fourier modes) with amplitudes $A_i$. To make contact with observations, we will restrict ourselves to a finite number $n$ of such modes across all fields. Thus our symplectic structure can be expressed in terms of functions $f_j$ of the amplitudes:

\be \omega = \sum_j f_j(\vec{A}) d\P_j \wedge dA_j \ee
where $\P_j$ is the conjugate momentum to $A_j$. We form the Liouville measure by raising the symplectic structure to a high enough power to cover the space of physical trajectories on a given Hubble slice:

\be \label{Liouville} \Omega = \omega^{n} \ee

Of particular importance for our analysis is the background volume mode $v$, whose contribution to the symplectic structure is $dv \wedge dH$. This mode itself is not an observable, but rather a gauge choice. In order to measure any physical length scale within cosmology, a reference scale must be given. A `rods and clocks' construction of cosmology must concede that the rods and clocks used are present in the evolving universe. Therefore there will exist a gauge redundancy under $v \rightarrow \lambda v$. One may choose a length scale in terms of the Fourier modes, but all that can be observed is the ratio of the wavelength of the mode to the total volume $v$, and thus an indistinguishable description of the same physical system will exist with a rescaled volume. It should be unsurprising to learn that the same physical system can be described in either metric or imperial units - although a physical length such as the volume of a closed universe will remain a physical observable, the choice of rod by which it is measured, be it the metre or foot, is a redundancy. This is reflected in the Hamiltonain constraint which can be expressed
\be 0= \H = v F[H, A_j, \dot{A_j}] \ee
and all physical observables at a given Hubble slice are composed of $A_j, \dot{A_j}$. Since a factor of $v$ can be pulled out of all observables, we find $\P_j = v f_j[\vec{A}, \vec{\dot{A}}]$ for some functions $f_j$. We require that the Hamiltonian constraint is satisfied, and so not all the $A_j$ and $\P_j$ are independent. We can therefore eliminate one of these using the constraint, and thus (up to an arbitrary choice of sign) the Liouville measure becomes 
\be \Omega=\dot{A}_o dv \wedge dA_o \bigwedge_{i \neq o} d\P_i \wedge dA_i \ee
and we can use the constraint to express $\dot{A}_o$ in terms of the remaining modes and the Hubble parameter. Rewriting this measure in terms of the $2n-1$-form $\Phi$ on the remaining phase space directions, we can express the Liouville measure as 
\be \Omega = M[H, \vec{A}, \vec{\dot{A}}] d\Phi \wedge dv \ee
wherein $M$ is the expression of $\dot{A_o}$ in terms of the remaining physical observables. 
Let us now construct the probabiltiy of observing an event $\E$. We suppose that the physical parameters in the early universe are distributed on an initial Hubble slice $H_1$ according to some unknown probability distribution $p$, thus the probability of observing $\E$ is
\be P(\E) = \int_R p[H,\vec{A}, \vec{\dot{A}}] d\Phi \ee
in which $R$ is the region of physical parameters for which the event $\E$ occurs. We are free to `thicken' this region by an interval $I$ in volume, which we shall take to be of unit length, to obtain an integral that is proportional to the Liouville measure (dropping the parameters from $p$ and $M$):
\be P(\E) = \int_{R \times I} p d\Phi \wedge dv = \int_{R \times I} \f{p}{M} \Omega \ee
An observer wanting to recreate this probability by making observations at a later time, and thus lower mean curvature $H'$ can do so defining R' to be the region to which trajectories in $R$ have evolved:
\be P(\E) = \int_{R'} p'[H',\vec{A}, \vec{\dot{A}}] d\Phi \ee 
To obtain $p'$ from $p$, note that Liouville measure is conserved, therefore we can again integrate this over a range $I'$ of the volume direction to eliminate the gauge direction, and recover a measure on the physical parameters. However, although $v$ is gauge in terms of observables, the relative expansion in the gauge direction is dependent on the physical parameters: To cover the same set as trajectories as was in $R \times I$ at the initial surface, we must allow $I'$ to vary. The trajectories lying in the interval $[v_1, v_2]$ on $H_1$ now occupy the range $[\lambda v_1, \lambda v_2]$ with $\lambda$ the ratio of volumes between slices:
\ba \label{expansion} \lambda &=& \exp[3 \int_{H_{1}}^{H_{2}} \f{H dH}{\dot{H}}] \nonumber \\
 		 &=& \exp[2 \int_{H_{1}}^{H_{2}} \f{dH}{H(1+w[H])}] \ea
in which we have expressed the ratio between the homogeneous components of pressure and energy density as $w=P/\rho$, for the effective Friedmann equation describing the evolution of the homogeneous modes \footnote{If we allow for phantom matter, in which $w < -1$ formula \ref{expansion} becomes singular. This is because the Hubble parameter is no longer necessarily monotonic. However, reparametrization by a different choice of clock will lead to a qualitatively similar result, with the resulting probability again dominated by maximally expanding solutions, in these cases dominated by the phantom matter.}. In eq (\ref{expansion}) the first line is generic across all gravitational theories as it represents a geometric identity, whereas the second represents the application of GR. The precise formulation of $w[H]$ is dependent upon the matter model and initial conditions, but this formulation holds for all weakly coupled matter models obeying the weak energy condition. Thus to recover $P(\E)$, we must integrate the Liouville measure over a range in volume. Since the $2n$-form separates into $dv \wedge d\Phi$, the volume integral can be performed first. Thus to obtain the probability of finding physical observables in state $X$ at $H_1$, which evolve to $X'$ and $H_2$ according to a probability distribution at high energy density, we must multiply the low energy distribution by the volume expansion:
\be \label{induced} p'(X) = \lambda(X') p(X') \ee
and so the probability must be weighted by the expansion in volume observed. It is easy to see from eq. (\ref{expansion}) that those solutions which have $w \approx -1$ for a large period of their evolution must be highest weighted, and that this weighting can be very large. In toy inflationary models consisting of a scalar field with a quadratic potential ($m=1.21 \times 10^{-6} M_{pl}$), taking the initial Hubble slice to be at the Planck scale this ratio reaches $ \exp[10^{13}] $. Thus the weighting can be completely overwhelming relative to any other measure considerations. If one pushes the initial distribution all the way back to the singularity in GR, to recover the weighting becomes a delta function on those trajectories which have maximal expansion. This is of course a purely mathematical exercise, as dimensional analysis informs us that quantum gravity effects must occur around the Planck regime, and thus our classical model is invalid here.

Anisotropies contribute a term that falls off as $v^{-2}$ (equivalently $a^{-6}$ for scale factor). This can be seen directly from the Bianchi I metric which describes the background modes of an anisotropic, homogeneous, flat cosmology in terms of the anisotropy parameters $\beta$ and $\sigma$. The spatial part of the metric has line element:
\be ds_3^2 = v^{2/3} \left( e^{-2\sigma}dx^2 + e^{\sigma-\f{\beta}{\sqrt{3}}} dy^2 + e^{\sigma+\f{\beta}{\sqrt{3}}}dz^2\right) \ee
When inserted into our action (\ref{action}) $\beta$ and $\sigma$ have the same form as massless scalar fields. Note that in this action, the values of $\sigma$ and $\beta$ are gauge, as a rescaling of coordinates can change them arbitrarily, however their kinetic terms are observable and contribute to dynamics. Again we must integrate the Liouville measure over a range in these gauge directions to project back down to a measure on physical observables. In this case, however, the interval is preserved under evolution, and so this is a trivial move. Those solutions which have high values of $\dot{\beta}$ and $\dot{\sigma}$ undergo far less expansion than more isotropic solutions, and are therefore disfavored after the effect of expansion in the volume gauge direction is considered. This provides a counter to the claim that isotropic universes are a set of measure zero \cite{Collins:1972tf} - as one takes the infinite limit of initial energy slice the distribution at high energy becomes a delta function on isotropy. 

A similar argument holds for inhomogeneities: Inhomogeneities distributed (as Fourier modes) at an initially high energy slice would lead to less expansion between surfaces of constant mean curvature. Therefore on examination of the distribution of anisotropies at low density, smaller amplitudes of inhomogeneous modes would be more strongly weighted to recover the high density distribution. This can be seen by considering the space of perturbative modes about a homogeneous (FRW or Bianchi I) background, but such an analysis ignores backreaction. The result holds for a constant mean curvature slicing in generality, and is therefore far stronger. This can be further generalized by substituting any monotonic observable (e.g. massless scalar field) to play the role of clock in situations where the mean curvature is not suitable. 

Volume weighted measures are often criticized for encountering the ``Q-catastrophe", in which it is noted that maximal expansion disfavours the amplitude of density fluctuations, $Q$, being the observed $10^{-5}$\cite{Qcat1}. This value is in the region considered to be necessary for the creation of galaxies in the era before dark energy domination, and is correlated with the flatness of any inflationary potential, with the formation of galaxies stopping if $Q<10^{-6}$, and failing due to over density if $Q>10^{-4}$ \cite{Tegmark:1997in}. The standard analysis which leads to the `catastrophe' is to assume an approximately flat prior on the mass (or other coupling constant) of the scalar field which is assumed to be the inflaton, and analyse the resulting the resulting distribution of $Q$, leading to a strong preference for large values of $Q$. However, it has been noted that a flat prior is not justified for this parameter as zero is a special point of the distribution, and in \cite{Garriga:2005ee}, a more refined distribution is explored. Here it is noted that the problem is inverted - such an approach leads to lower values of $Q$ being preferred, with $10^{-6}$ being orders of magnitude more likely than the observed $10^{-5}$. Such issues are avoided in the multiverse context by the application of a scale factor cut-off \cite{DeSimone:2008bq}, however such a cut-off is incompatible with evolution of the Liouville measure - different initial conditions will cross any cut-off at a rate dependent on their physical parameters. A separate solution is offered through decay of the inflaton \cite{Hall:2006ff} or through a specific choice of landscape. 

In such analyses, it is assumed that inflation is due to a scalar field subject to a power-law potential, and that this field is also responsible for the perturbation spectrum that is observed. There exist a whole plethora of possible inflationary mechanisms, and the set in which this is due to a scalar field subject to power law potential are but a small fraction of these. Furthermore, highly successful models such as the curvaton \cite{Lyth:2002my} and multi-field inflationary models \cite{Easther:2013rva} are claimed not to suffer from this issue at all: In such models the amplitude of fluctuations is decoupled from the behaviour of the inflaton, removing the Q-catastrophe. However such models require that the mass of the curvaton, for example, is forced far below that of the inflaton, and thus the decoupling may not be complete \footnote{We are grateful to the referee who pointed out this subtle issue.}. We note at this point that the Planck data indicate that such single field models are observationally disfavored \cite{Ade:2015lrj}, and therefore it is natural to consider modified scenarios such as multifield models. The question of the Q-catastrophe, and its possible resolution, rests on probability distributions on the landscape, which we do not address. The analysis that we present here makes no claim on the nature of the inflaton - the re-weighting of measures by volume due to the conservation of the Liouville form is independent of the matter playing the role of inflaton. This can even be extended to modified gravitational actions in which the inflationary behaviour is due to higher order terms of the Ricci scalar (for example) for which there is no Q-catastrophe, as all that is required is the conservation of the symplectic form, a standard result of Hamlitonian physics. 

The idea of generality is of course vague\cite{Barrow:2015fga} and often subject to the biases of the observer's preferred set of variables \cite{Norton}: One cannot define a probability distribution at high energy without reference to some more fundamental theory. There are several alternative approaches to defining cosmological probabilites, particularly in the areas of inflation. Most of these, however, start from a model which involves directly the gauge direction in volume, leading to a host of measure issues\cite{Vilenkin:1994ua}. These arise most acutely in the context of eternal inflation, in which a (countably) infinite number of ``bubble universes" can form, and thus probability definitions unavoidably encounter classical problems of counting infinite sets in which any probability that is desired can be obtained by re-ordering the set. Attempts to solve this often involve ad-hoc arguments, such as cut-offs in time \cite{Guth:2011ie} or those involving maximizing numbers of observers \cite{Vilenkin:1994ua}. 

However, these arguments all have significant flaws: There is no way for an observer in a post-eternal inflationary universe to know how long the universe spent in this phase, and indeed if the nucleation rate is high enough, for any given time $T$ the probability that we spent longer than $T$ in this state approaches 1 as the cut-off time tends to infinity. Maximizing observers is also heavily dependent on the order of counting - it is ill-defined to ask whether it is more probable that we belong to one of the infinite number of universes with a single observer, or a single universe with an infinite number? The argument put forth here is that one should be agnostic on such issues, as they represent unobservable parameters. The result presented is that the manner in which initial parameters for a post-quantum inflationary phase are distributed is largely irrelevant. What matters is not what die is cast, nor how it is cast, but rather when: by whatever means the physical parameters are distributed at some initial energy density, the probability that an observer at a lower energy density observes an event is weighted by the relative expansion in volume of the set of universes for which that event occurs. This weighting favours heavily those solutions which underwent a large expansion, i.e. for matter obeying the weak energy condition, those which had an extended period in which $w=-1$, and are closest to isotropy and homogeneity, as these experience the greatest expansion. Note that the purely isotropic, homogeneous, inflation-dominated universe is a set of measure zero on the space of all possible trajectories since these form a continuum, so a typical observer should expect to find conditions in a small neighborhood around this point, just as an exponential distribution is peaked at zero - this is the most probable state, but any measurement should be expected to be in a region close to this. 

This inverts the usual claim of fine-tuning of initial parameters: it would in fact require that at the point of inception, the probability distribution of physical parameters would have to be heavily fine-tuned to avoid a large, isotropic universe which experienced slow-roll inflation. Maximal expansion can of course be generated by a pure cosmological constant. However, since we are interested in the observations made at low energy density (such as in the universe that we currently inhabit) this precludes the case of $\rho_\Lambda$ exceeding this value. Such reasoning is, of course, anthropic and follows the arguments of Weinberg \cite{Weinberg:1987dv} - volume weighting of probabilities does not explain why the cosmological constant is within any set of bounds, but strongly favours the maximization of the cosmological constant subject to this condition. Thus our analysis does not explain the existence of such a bound, but rather its saturation. A low energy observer should expect to see a universe dominated by conditions very close to $w=-1$ with $\Omega = 1$. 

Our derivation of volume weighting applies during the classical, slow-roll expansion of the universe regardless of any eternal inflationary phase. This derivation complements the arguments of Linde et al. \cite{Linde:2007nm, Linde:2008xf} who proposed volume weighting during slow roll in the context of eternal inflation. We have shown that distributions at high energy density are re-weighted by the flow of the Liouville measure under Hamiltonian evolution. Given an distribution of physical parameters from some pre slow-roll phase (be it based on eternal inflation, bouncing cosmologies or any other considerations) our analysis shows that on making low energy observations after slow-roll, the conservation of the symplectic structure introduces naturally a volume weighting. 

This publication was made possible through the support of a grant from the John Templeton Foundation. The authors are very grateful to the anonymous referee whose input has greatly improved the work. 

\bibliographystyle{unsrt}
\bibliography{letter}

\end{document}